\documentclass[preprint,12pt]{elsarticle} 
\usepackage{booktabs}
\usepackage{amssymb}
\usepackage[colorlinks]{hyperref}   
\usepackage{xcolor}
\usepackage{amsthm}
\usepackage{mathtools}
\usepackage{enumitem}
\usepackage{graphicx}
\usepackage{subfigure}
\usepackage{autobreak}
\usepackage{lineno}
\usepackage{natbib}
\usepackage[margin=2cm]{geometry}
\usepackage{multirow}
\usepackage{floatrow}
\usepackage{colortbl}
\usepackage{threeparttable}

\usepackage{graphicx}
\newfloatcommand{capbtabbox}{table}[][\FBwidth]

\setcitestyle{authoryear,open={(},close={)}}
\definecolor{orange}{rgb}{1,0.5,0}
\definecolor{red}{RGB}{198,0,35}
\definecolor{amberseldef}{rgb}{1.0, 0.49, 0.0}
\definecolor{ceruleanblue}{rgb}{0.16, 0.32, 0.75}
\definecolor{amber}{rgb}{1.0, 0.49, 0.0}
\definecolor{dodgerblue}{rgb}{0.12, 0.56, 1.0}
\definecolor{pureblue}{rgb}{0, 0, 1.0}

\definecolor{blue}{rgb}{0.0, 0.28, 0.67}

\def\hmath$#1${\texorpdfstring{{\rmfamily\textit{#1}}}{#1}}

\makeatletter
\def\ps@pprintTitle{%
   \let\@oddhead\@empty
   \let\@evenhead\@empty
   \let\@oddfoot\@empty
   \let\@evenfoot\@oddfoot
}
\makeatother

\AtBeginDocument{\hypersetup{citecolor= pureblue,linkcolor = pureblue,urlcolor = dodgerblue}}
\begin{document}


\begin{frontmatter}


\address[1]{Department of Civil Engineering, the University of Hong Kong, Hong Kong, China}
\address[2]{Department of Urban and Rural Planning, School of Architecture and Design, Southwest Jiaotong University, Chengdu, China}
\cortext[cor1]{Corresponding author: kejintao@hku.hk}

\author[1]{Wang Chen}
\author[1]{Jintao Ke \texorpdfstring{\corref{cor1}}{}}
\author[2]{Linchuan Yang}

\title{Scaling Laws of Dynamic High-Capacity Ride-Sharing}

\begin{abstract}

Dynamic ride-sharing services, including ride-pooling offered by ride-hailing platforms and demand-responsive buses, have become an essential part of urban mobility systems. These services cater to personalized and on-demand mobility requirements while simultaneously improving efficiency and sustainability by accommodating several trip requests within a single ride. However, quantifying the advantages and disadvantages of dynamic ride-sharing, particularly high-capacity ride-sharing, remains a challenge due to the complex dynamics that depend on several factors, including matching algorithms, vehicle capacity, transportation network topology, and spatiotemporal demand and supply distribution. In this study, we conduct extensive experiments on an agent-based simulation platform calibrated by real-world mobility data from Chengdu, Hong Kong, and Manhattan. Our findings reveal a few scaling laws that can effectively measure how key performance metrics such as passenger service rate and vehicle occupancy rate change with a dimensionless system loading factor that reflects the relative magnitude of demand versus supply. Moreover, our results indicate that these scaling laws are universal for different network topologies and supply-demand situations. As a result, these scaling laws offer a means for urban planners, city managers, and ride-hailing platforms to quantify the potential benefits and drawbacks of dynamic ride-sharing under different circumstances and to design better operational and regulatory strategies.

\end{abstract}


\end{frontmatter}


\section{Introduction}\label{sec: 1}

Human mobility has been essential to urban life for over 7,000 years \citep{mumford1961city}. However, urban human mobility is a complex system that involves the interactions of individual travelers with each other, the environment, and different transportation services \citep{schroder2020anomalous, barbosa2018human}. Urban human mobility needs are primarily satisfied by public transportation (metro, bus), taxi, and private car \citep{vuchic2017urban}. While public transportation services are space-efficient (with each passenger occupying less road space), they can only accommodate for a fraction of travel demand due to fixed routes and limited coverage \citep{vuchic2017urban}. As an essential complement to public transport, private vehicles and taxis can cater for door-to-door and on-demand travel needs at 24-h-a-day availability. However, the vehicle occupancy rate of taxis and private vehicles is generally low with on average 1.3 passengers in each vehicle \citep{mitchell2010reinventing}, causing traffic congestion and carbon emissions \citep{erhardt2019transportation, arnott1994economics}. With the proliferation of smart phones and mobile internet technologies, on-demand ride-sharing services have experienced rapid growth in recent years. By using one vehicle to serve two or more passengers in each ride, these services can improve space-efficiency and vehicle utilization while maintaining the flexibility of taxis and private vehicles \citep{mcdonnell2016ecological, storch2021incentive, un2015world}.

Dynamic ride-sharing systems provide door-to-door services for a few passengers with similar itineraries and schedules and enables them to split the costs. Ride-sharing programs offer numerous benefits to drivers and passengers, such as saving travel costs and increasing vehicle utilization, while also benefiting society by enhancing mobility, alleviating traffic congestion, and reducing carbon emissions and pollution \citep{chan2012ridesharing, santi2014quantifying,furuhata2013ridesharing}. The first ride-sharing program dates back to World War \uppercase\expandafter{\romannumeral2}, when the US government established Car-sharing Clubs to conserve resources for the war \citep{chan2012ridesharing}. Traditional ride-sharing programs, including carpooling, vanpooling, and dial-a-ride, require prearrangement, with users submitting their trip schedules and itineraries in advance to an intermediate platform that matches passenger requests. However, with the proliferation of smartphones and mobile internet, dynamic ride-sharing programs that can match drivers and passengers on very short notice or even en route have become increasingly popular worldwide \citep{shaheen2015shared, agatz2012optimization, chen2017understanding}. 

Dynamic ride-sharing has become an important service option offered by many ride-hailing platforms, such as Uber, Lyft, and Didi. These companies have strong incentives to enhance vehicle utilization due to supply shortages \citep{ke2020pricing}. Recently, some public transit operators introduce demand-responsive buses, which can also be regarded as a type of high-capacity dynamic ride-sharing, as the routes of these buses can be dynamically adjusted to visit heterogeneous origins and destinations of various passengers in a single ride \citep{huang2020two, vansteenwegen2022survey}. While there is a general consensus that ride-sharing can improve vehicle utilization and traffic efficiency, little is known about how and to what extent these metrics are affected by ride-sharing with different vehicle capacities and under different circumstances. This information is critical for both operators (such as ride-hailing platforms and bus companies) and governments, as it can help them make better decisions in operations and regulations. For example, when the system is heavily loaded, meaning that there is an excessive demand relative to supply, high-capacity ride-sharing programs are more desirable since each vehicle can serve more passengers in each ride, leading to a higher order fulfillment rate and lower passenger waiting time. In contrast, when the system is lightly loaded, that is, demand is relatively small compared to supply, introducing high-capacity ride-sharing could be detrimental as it contributes little to improving vehicle utilization but may cause longer detour times to passengers.

A few recent works have been devoted to this critical problem. Unlike normal ride-hailing services that assign a passenger to a single driver, a dynamic ride-sharing system needs to assign multiple passengers to a driver who is either idle or already on route for delivering other passengers, leading to a much more complex matching process and a much larger solution space for seeking an optimal assignment \citep{santi2014quantifying, alonso2017demand}. \cite{santi2014quantifying} proposed shareability networks to efficiently solve the assignment problem for dynamic ride-sharing. They conducted a series of simulation studies in Manhattan to to quantify the benefits of ride-sharing, including saved travel time, saved trips, and improved vehicle utilization. In real practice, however, it is usually significantly time-consuming to run extensive simulations, thus concise formulas or empirical laws for approximating the performance metrics of dynamic ride-sharing services are warranted. In addition, a few studies have analysed various benefits of ride-sharing services compared to ride-hailing services in different cities \citep{narayan2022scalability, soza2022shareability, liu2023scale}. However, these studies did not also explicitly formulate the implications of ride-sharing on system performances under different circumstances (e.g., different demand-to-supply ratios). To address this issue, researchers \citep{tachet2017scaling, ke2021data} proposed a few scaling laws to quantify the sharing probability against passenger demand based on the simulation results in different cities. Some mathematical models have also been proposed to calculate the sharing probability \citep{daganzo2020analysis, wang2021predicting}. However, these studies focus on matching at most two passengers and cannot be adapted to high-capacity ride-sharing scenarios (i.e., three or more passengers share one vehicle) due to the difficulties in estimating routing distance for complex routes that visit multiple origins and destinations. Combining numerical simulation and a mean-field analysis, \cite{molkenthin2020scaling} estimated the functional form of a scaling law in a simplified dynamical model of ride-sharing in the special case of unlimited vehicle capacity. Moreover, \cite{zech2022collective} relaxed the strong assumption of unlimited vehicle capacity and analyzed the dynamics of ride-sharing under the vehicle capacity constraint on different network topologies, e.g., two-dimensional square lattices. Nevertheless, their proposed scaling laws cannot estimate many key performance metrics of dynamic ride-sharing, such as passenger service rate, and how these metrics are influenced by the relatively magnitude of supply and demand.

In response to these challenges, this study makes one of the first attempts to quantify the efficiency of dynamic high-capacity ride-sharing services under different supply-demand conditions and on various network topologies. Through the experiments carried out on an agent-based simulation platform calibrated using real-world datasets from Chengdu, Hong Kong, and Manhattan, we have identified serveral concise scaling laws that can well characterize the trends of key system performance measures such as vehicle occupancy rate and passenger service rate against a dimensionless parameter called system load. Our findings show that these scaling laws can universally approximate these relationships with reasonably high goodness-of-fit across different supply-demand patterns, vehicle capacities, and network topologies. These scaling laws provide critical insights into the performance of high-capacity ride-sharing services and can assist urban planners, city managers, and ride-hailing platforms in making informed decisions regarding operational and regulatory strategies under different circumstances.


\section{Results}\label{sec: 2}
\subsection{System load}\label{sec: 2.1}
Consider a dynamic ride-sharing system with a vehicle fleet size of $N$, an average velocity of vehicles of $v$, an arrival rate of passenger requests of $\lambda$, and an average in-vehicle travel distance of all ride requests of $\Bar{d}$. In scenarios without ride-sharing, the average in-vehicle travel time of all ride requests is $\Bar{d}/v$. Unlike traditional street-hailing taxi systems in which a matching occurs only when a passenger and a driver physically meet each other at a specific spot, a dynamic ride-sharing system allows an online matching between passengers and drivers within a certain distance. The time a driver takes to pick up their assigned passenger is referred to as the average pickup time, while the average service time of drivers, denoted by $\Bar{t}$, consists of pickup time and in-vehicle time in a ride. When ride-sharing is applied, a driver is assigned to pick up and deliver multiple passengers in each ride, adding an extra detour time to the in-vehicle travel time. Due to the pickup time and extra detour time for ride-sharing, the average service time of drivers in one ride $\Bar{t}$ is generally larger than $\Bar{d}/v$, which can be regarded as the lower bound of $\Bar{t}$ being achieved in an ideal situation without pickup and detours.

The average service time $\Bar{t}$ depends on the matching algorithms (regarding matching radius and maximal allowable detour time), road network topology, and maximal capacity of vehicles. For instance, a larger matching radius allows for the possibility of picking up passengers by more distant drivers, resulting in longer average pickup times and thus a larger $\Bar{t}$. Meanwhile, the average detour time of passengers as well as $\Bar{t}$ will be longer under situations with a longer maximal allowable detour time, a larger vehicle capacity, or a more complex road network topology. At equilibrium, $\lambda$ and $N/\Bar{t}$ respectively represent the arrival rate of passengers and drivers, namely, the number of passengers and drivers arriving at the market per unit time. Now we introduce a dimensionless parameter named the system load to reflect the demand-to-supply ratio of a dynamic ride-sharing system, as follows:

\begin{equation}
    \label{eq:load}
    u = \frac{\lambda}{N/\Bar{t}}
\end{equation}

This system load $u$ represents the relative magnitude of demand versus supply in a dynamic ride-sharing system. In addition, $u$ contains the average service time $\Bar{t}$ that depends on the matching algorithm, road network topology, and maximal capacity of vehicles. As such, the system load $u$ has the potential to capture all key characteristics of a dynamic ride-sharing system, including the demand-supply patterns, matching algorithms,traffic road network, and maximal capacities of vehicles. Ideally, when $u \leq 1$, i.e., the number of required trips per unit time is no more than the number of idle vehicles per unit time, the platform can assign each driver to serve one passenger at a time without using ride-sharing. When $u > 1$, however, ride-sharing must be utilized to meet the excess demand.

\subsection{Scaling laws}\label{sec: 2.2}

\begin{figure}[!h]
    \centering
    \subfigure[Chengdu]{\includegraphics[height = 0.25\linewidth, width=1.0\linewidth]{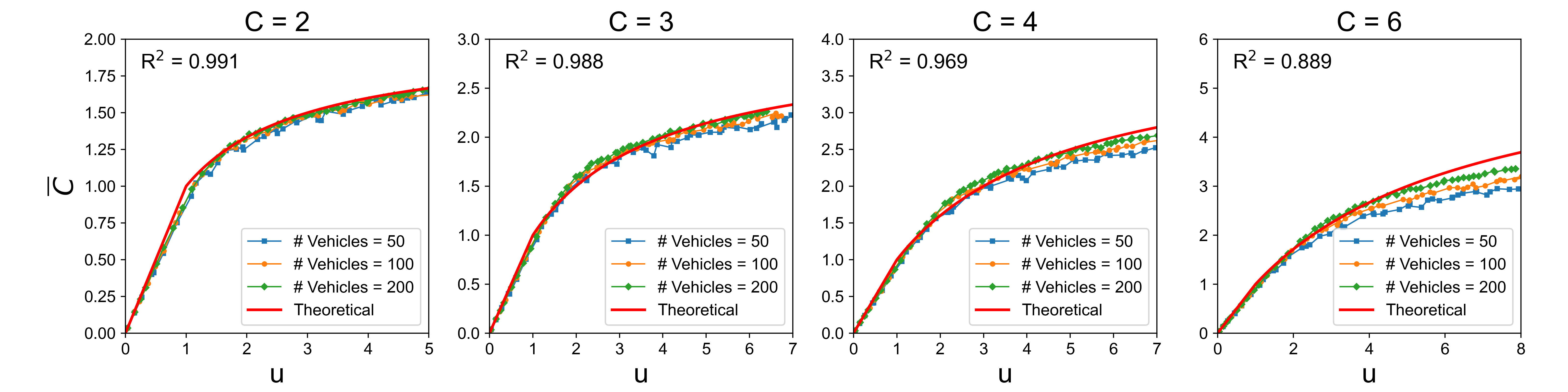}}
    \subfigure[Hong Kong]{\includegraphics[height = 0.25\linewidth, width=1.0\linewidth]{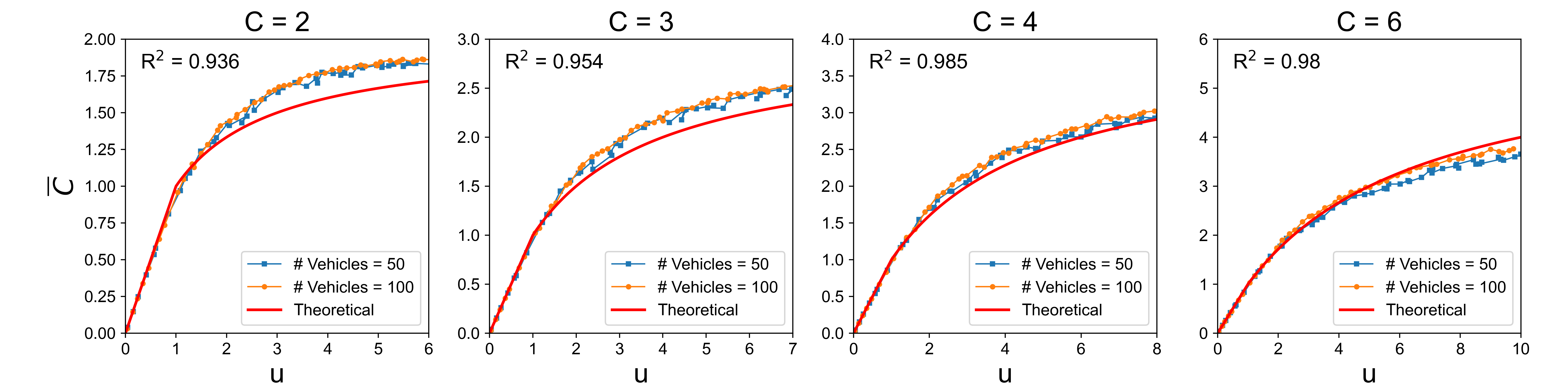}}
    \subfigure[Manhattan]{\includegraphics[height = 0.25\linewidth, width=1.0\linewidth]{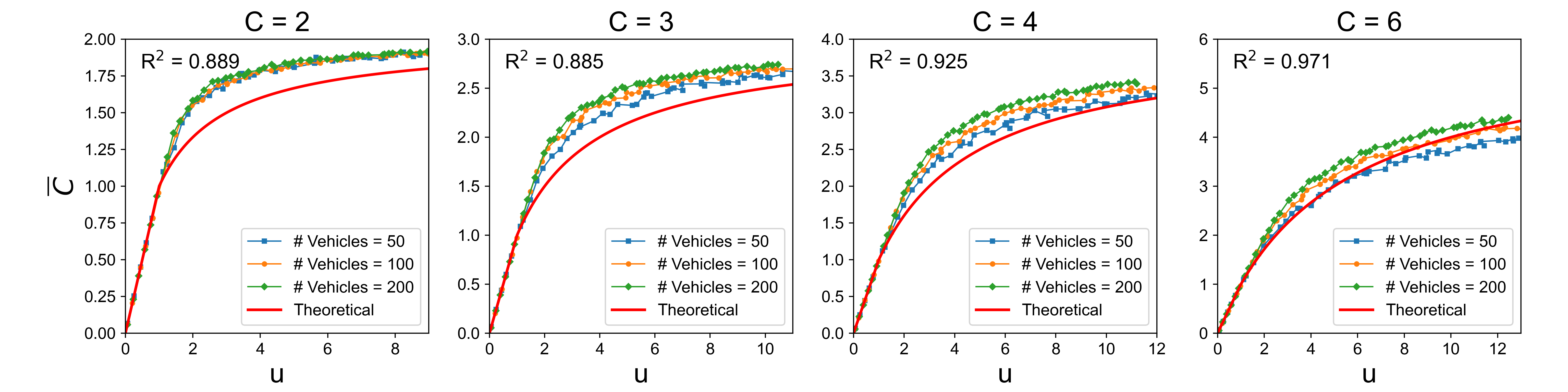}}
    \caption{\textbf{Scaling of the average number of scheduled passengers $\Bar{C}$.} Prediction of the theoretical model is in great agreement with experimental results. The maximal fleet size is 100 for Hong Kong due to limited demand. The minimal $R^2$ across all scenarios is 0.885. See Table \ref{table1} for more details.}
    \label{fig1}
\end{figure}

\begin{figure}[!h]
    \centering
    \subfigure[Chengdu]{\includegraphics[height = 0.25\linewidth, width=1.0\linewidth]{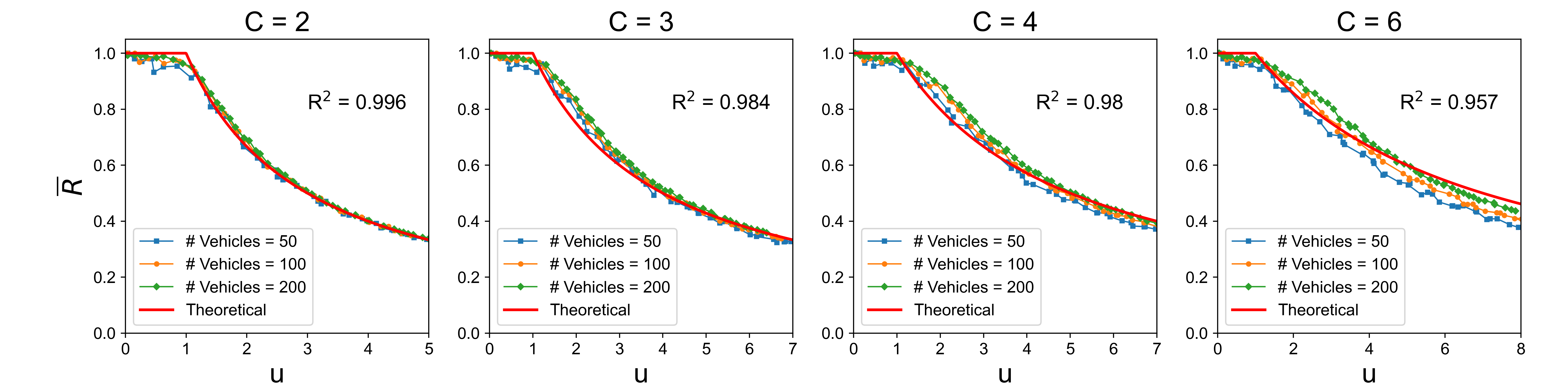}}
    \subfigure[Hong Kong]{\includegraphics[height = 0.25\linewidth, width=1.0\linewidth]{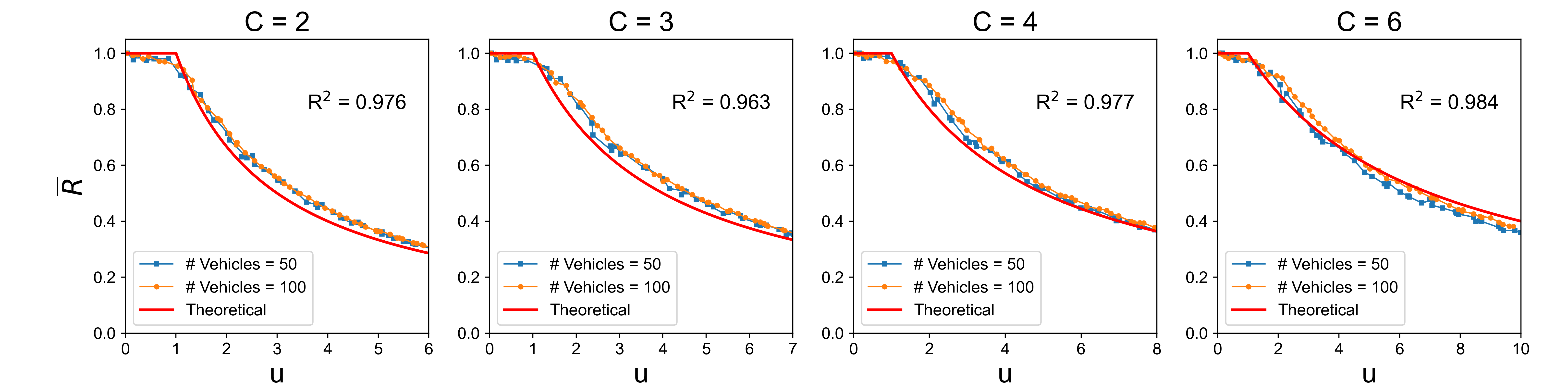}}
    \subfigure[Manhattan]{\includegraphics[height = 0.25\linewidth, width=1.0\linewidth]{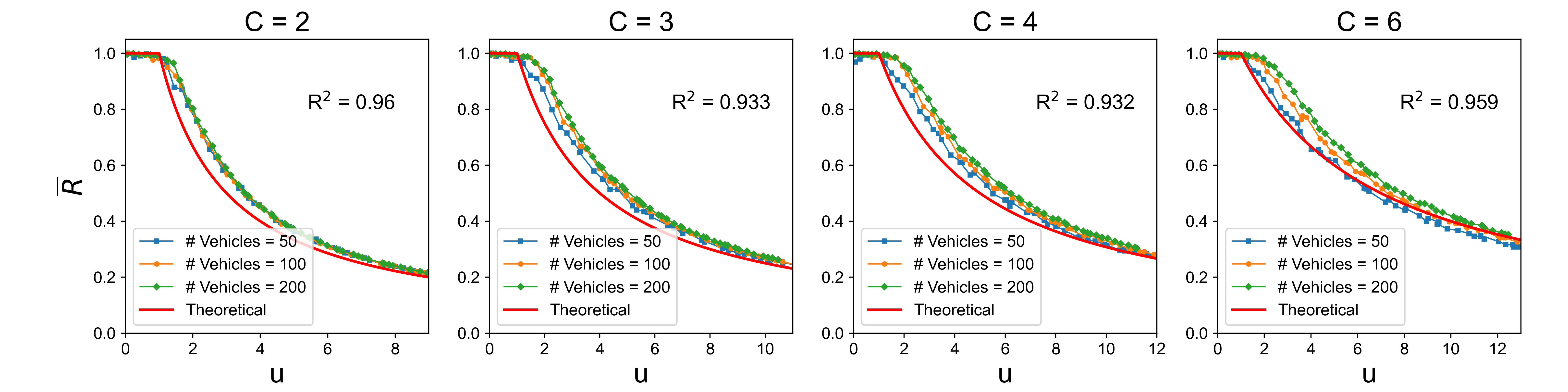}}
    \caption{\textbf{Scaling of the average passenger service rate $\Bar{R}$.} Prediction of theoretical model is in great agreement with experimental results. The minimal $R^2$ across all scenarios is 0.932. See Table \ref{table2} for more details.}
    \label{fig2}
\end{figure}

Since the system load $u$ can capture the key features of a dynamic ride-sharing system, it is of immense interest to explore how the performances of a dynamic ride-sharing system change with this system load $u$. We conduct extensive experiments on an agent-based simulation platform calibrated using real request data and road networks in Chengdu, Hong Kong, and Manhattan, and investigate the trends of vehicle occupancy rate $\Bar{C}$ and passenger service rate $\Bar{R}$ against the system load $u$. Specifically, vehicle occupancy rate $\Bar{C}$ refers to the average number of scheduled passengers including passengers already on the vehicle and passengers scheduled to be picked up in the future. Passenger service rate $\Bar{R}$ refers to the ratio of the number of served requests to the total number of requests. Figs. \ref{fig1} and \ref{fig2} respectively illustrate the trends of $\Bar{C}$ and $\Bar{R}$ with respect to $u$ in different cities. The experimental results show that $\Bar{C}$ monotonically increases with $u$, while $\Bar{R}$ monotonically decreases with $u$. Furthermore, we observe $\Bar{C}$ and $\Bar{R}$ exhibit similar trends across different city network topologies, vehicle densities, and vehicle maximal capacities, suggesting that the performances of dynamic high-capacity ride-sharing (reflected by $\Bar{C}$ and $\Bar{R}$) can be characterized by scaling laws as a function of the system load $u$.

We now propose the following empirical scaling laws to approximate $\Bar{C}$ and $\Bar{R}$:

\begin{equation}
    \label{eq:C_bar}
    \Bar{C}=\left\{
        \begin{array}{cl}
            u        & u \leq 1 \\
            \frac{u}{C - 1 + u} \cdot C & u > 1    \\
        \end{array} \right.
\end{equation}

\begin{equation}
    \label{eq:R_bar}
    \Bar{R}=\left\{
        \begin{array}{cl}
            1         & u \leq 1 \\
            \frac{1}{C-1 + u} \cdot C & u > 1    \\
        \end{array} \right.
\end{equation}
where $C$ denotes the maximal capacity of vehicles. It is worth noting that \textbf{Eqs.~\eqref{eq:C_bar} and \eqref{eq:R_bar} have only one parameter $u$}, which means these empirical laws can concisely reveal the effects of system load $u$ on the key system performance metrics. As shown in Eqs.~\eqref{eq:C_bar} and \eqref{eq:R_bar}, as system load $u$ approaches 0, the average number of scheduled passengers $\Bar{C}$ tends to be 0, and the average service rate of passengers $\Bar{R}$ approaches 1. This is reasonable because when the passenger demand is extremely smaller than the vehicle supply, almost all vehicles are idle in the ride-sharing system, i.e, $\Bar{C} \rightarrow 0$. Hence, each passenger can be assigned to an idle vehicle nearby, resulting in all passenger requests being satisfied in the system, i.e., $\Bar{R} \rightarrow 1$. On the other hand, as $u$ approaches infinity, $\Bar{C}$ tends to be $C$, and $\Bar{R}$ tends to be 0. This is because when the passenger arrival rate is significantly higher than the driver arrival rate, there are many passengers but few idle vehicles in the system, leading to almost all requests cannot be satisfied, i.e., $\Bar{R} \rightarrow 0$. Meanwhile, when an idle vehicle becomes available in the system, the driver can serve $C$ passengers with similar origins and destinations since there are adequate passengers in the system, i.e., $\Bar{C} \rightarrow C$.

We then compare the proposed scaling laws with the experimental outcomes generated by the agent-based simulation platform calibrated by real-world mobility data from three cities (i.e., Chengdu, Hong Kong, and Manhattan). As shown in Figs. \ref{fig1} and \ref{fig2}, the scaling laws demonstrate a strong agreement with experimental results across different cities, with $R^2$ between 0.885 and 0.996. Further details on error measures under various conditions and in different cities are presented in Tables \ref{table1} and \ref{table2}.

\section{Discussion}\label{sec: 3}

The main contribution of this study is the identification of a few universal scaling laws that can measure the efficiency of dynamic high-capacity ride-sharing systems as a function of the system load that reflects the demand-to-supply ratio. These scaling laws are concise and simple, requiring only one parameter, the system load, and are applicable to scenarios involving different vehicle capacities, road network topologies, matching algorithms, and supply-demand patterns. Additionally, the system load can be easily approximated using a few predefined exogenous variables (see Approximation of $u$ in Supplementary Information for details), making the derived scaling laws valuable for predicting and measuring the performances of dynamic high-capacity ride-sharing systems in various cities with distinct characteristics.

Dynamic high-capacity ride-sharing is a complex system with intricate interactions among its exogenous and endogenous variables, making it challenging to develop a mathematical model that can accurately capture its performance. In this study, we present a novel approach by deriving concise scaling laws based on fundamental physical principles (refer to Derivation in Supplementary Information for additional details). Through extensive experimentation in three cities (Chengdu, Hong Kong, and Manhattan), we demonstrate the effectiveness of these scaling laws in approximating the system's performance.

As shown in Figs. \ref{fig1} and \ref{fig2}, the density of vehicles has little effect on the performance of dynamic ride-sharing when the system load $u$ is the same. This demonstrates that the proposed scaling laws are universal in the sense that the efficiency of dynamic ride-sharing (reflected by $\Bar{C}$ and $\Bar{R}$) can be mainly captured by the value of the system load $u$, and the trends are independent of other variables. Furthermore, the scaling laws for high-capacity ride-sharing services (i.e., C $\geq$ 4) approximate the experimental outcomes almost as accurately as those for low-capacity ride-sharing (i.e., C = 2). This indicates that these scaling laws are adaptable not only to low-capacity ride-sharing but also to high-capacity ride-sharing, which fills the research gap of existing scaling laws/theoretical models that primarily characterize ride-sharing programs with at most two passengers being pooled together in a ride.

To determine whether multiple requests can be pooled to a driver, we need to design a specific route for the driver to sequentially visit the origins and destinations of all requests. The optimal route, i.e., the route with the minimum cost, can be obtained by enumerating all possible routes that visit all origins and destinations and comparing their corresponding costs. If the optimal route satisfies all constraints of maximal pickup and detour time, the requests can be assigned to the driver. However, in high-capacity scenarios (e.g., C = 6), it is significantly time-consuming to enumerate all possible routes to check whether multiple trip requests can be shared due to the large number of possible routes. Thus, a heuristic algorithm, i.e., the nearest neighbor (NN) algorithm, is used to check shareability among multiple trip requests. According to the NN algorithm, the driver always visits the nearest spot (origin or destination) to pick up or deliver passengers. Thus, we can obtain the locally optimal route without enumerating all possible routes. The locally optimal route obtained through the NN algorithm typically has a higher cost than the optimal route, which may result in longer pickup or detour times that do not satisfy the constraints of the maximal pickup or detour time. Consequently, the sharing of multiple trip requests that could have been pooled to the driver may be denied. Nevertheless, the NN algorithm can still achieve acceptable accuracy with much lower computation costs \citep{rosenkrantz1977analysis}.

The average number of scheduled passengers $\Bar{C}$ is slightly underestimated by the scaling laws in Hong Kong and Manhattan. This could be due to the fact that the study areas in Hong Kong and Manhattan are relatively small, making it easier for passengers to share a vehicle since their origins and destinations are closer. However, when the vehicle maximal capacity is increased to 6 (i.e., C = 6), the errors between the experimental results and the proposed scaling laws are smaller. This is mainly because the NN algorithm used in the experiments may slightly underestimate the possibility of sharing among multiple trip requests, leading to fewer passengers being carried per vehicle on average and a lower service rate than expected. Nevertheless, the $R^2$ values for $\Bar{C}$ and $\Bar{R}$ in all scenarios are more than 0.885 and 0.932, respectively, and the mean absolute percentage errors (MAPEs) are no more than 10\% (Tables \ref{table1} and \ref{table2}). These satisfactory approximation results demonstrate the reasonability, effectiveness, applicability and universality of the proposed scaling laws.

The proposed scaling laws have significant implications for the operations, management, and regulations of on-demand high-capacity ride-sharing programs. Service operators and regulators can use these laws to gain a better understanding of the advantages and disadvantages of high-capacity ride-sharing under different supply and demand situations, allowing for the design of more efficient and sustainable urban mobility systems. Specifically, the average number of scheduled passengers $\Bar{C}$ indicates the occupancy rate of vehicles, which can further be used to estimate the probability of a passenger sharing a ride with others and the average revenue of drivers. These are key performance metrics that can impact the benefits/costs of different stakeholders, including platforms, drivers, and passengers, and their willingness to participate in the market. On the other hand, the average service rate of passengers $\Bar{R}$ reflects the efficiency of the ride-sharing transportation system, which can guide fleet size management so as to estimate the reduction of congestion and carbon emissions resulting from dynamic ride-sharing services. In particular, high-capacity ride-sharing programs can achieve the same passenger service rate with a smaller fleet size, reducing traffic congestion and mitigating carbon emissions. These empirical laws provide valuable insights into the selection of appropriate operational and regulatory strategies under different circumstances. Generally, when the demand-to-supply ratio is high, high-capacity ride-sharing programs are recommended because they can improve vehicle utilization, reduce vehicle-mile-traveled, mitigate carbon emissions, and serve more passengers. For instance, in Manhattan, when the system load $u$ is 4, high-capacity ride-sharing services (C = 6) achieve a service rate of approximately 75\%, whereas low-capacity ride-sharing services (C = 2) achieve a service rate of around 45\%. This demonstrates that high-capacity ride-sharing services can accommodate for significantly more passengers.

\section{Methods}\label{sec: 4}
\subsection{Datasets}\label{sec: 4.1}

\begin{figure}[!h]
    \centering
    \subfigure[]{\includegraphics[width=0.5\linewidth]{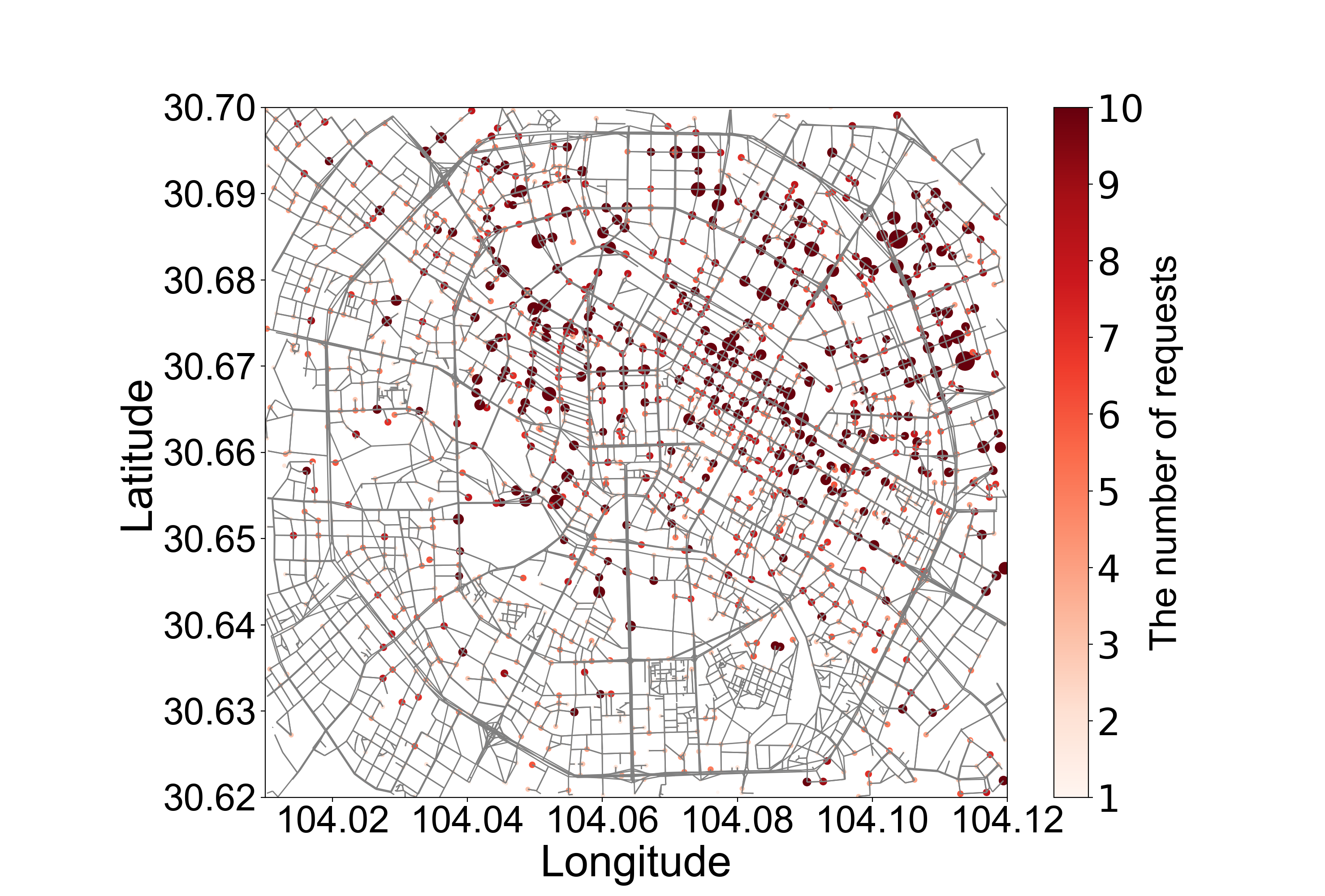}}
    \subfigure[]{\includegraphics[width=0.4\linewidth]{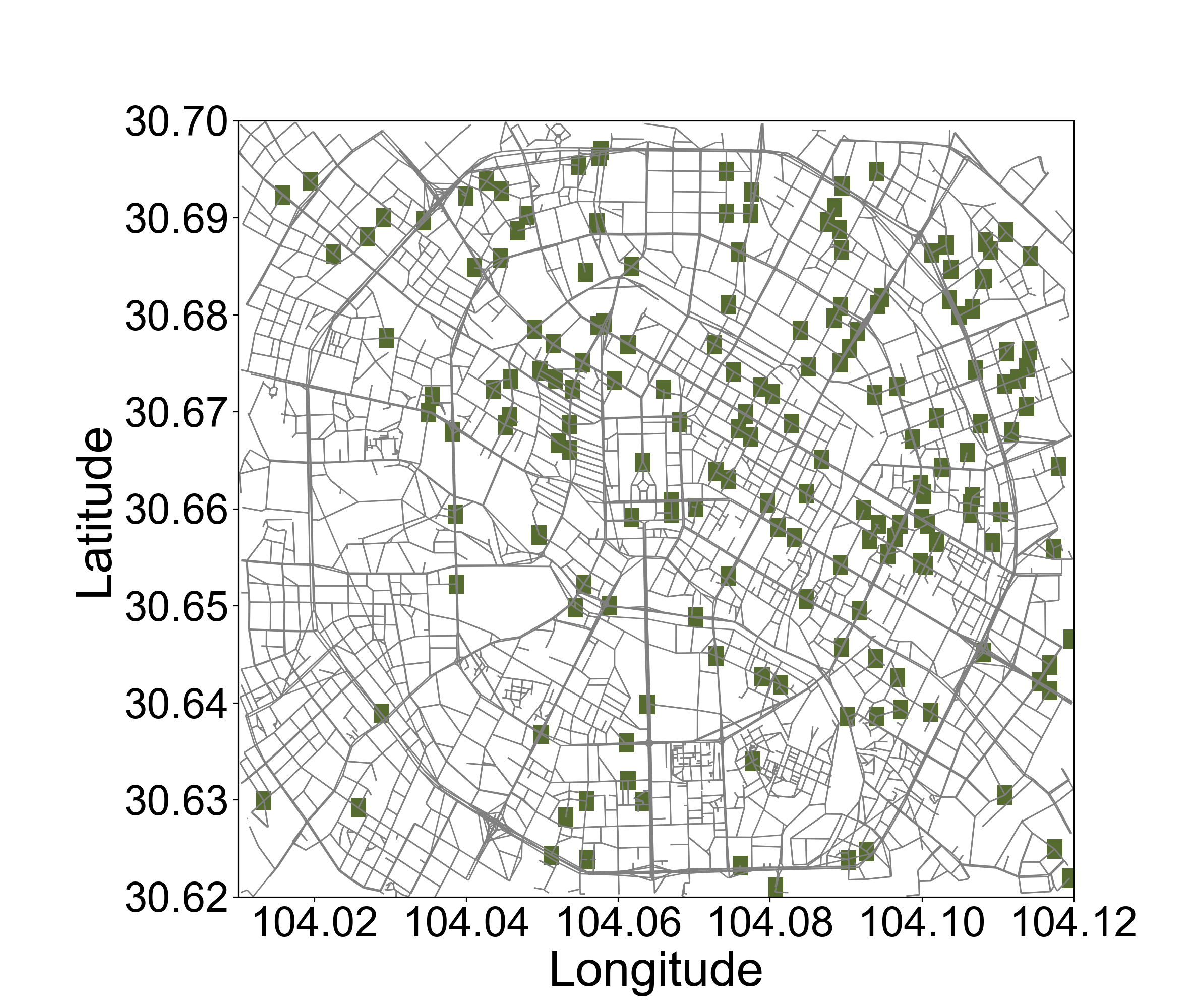}}
    \caption{\textbf{Distributions of requests and vehicles in Chengdu.} The northeastern area is a relatively hot zone and vehicles are initially concentrated in this zone. Red circles and green squares represent passengers and vehicles, respectively (similarly hereinafter). The request arrival rate in (a) is 0.76 \#/s, and the number of vehicles in (b) is 200.}
    \label{fig3}
\end{figure}

\begin{figure}[!h]
    \flushleft
    \subfigure[]{\includegraphics[width=1.0\linewidth]{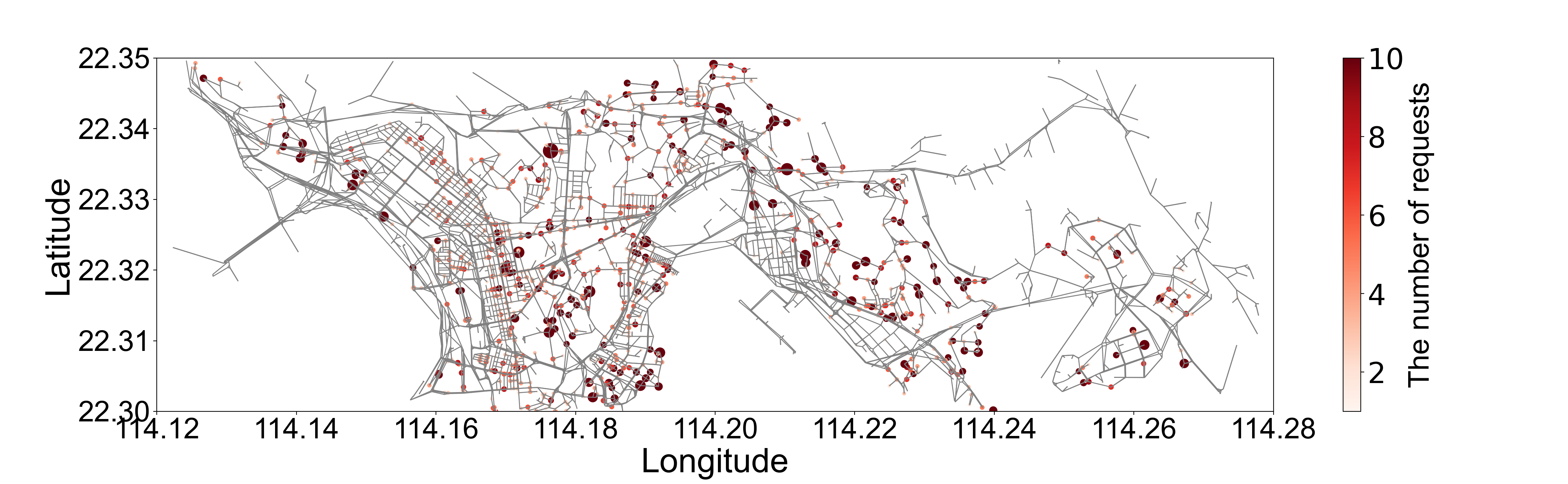}}
    \subfigure[]{\includegraphics[width=0.9\linewidth]{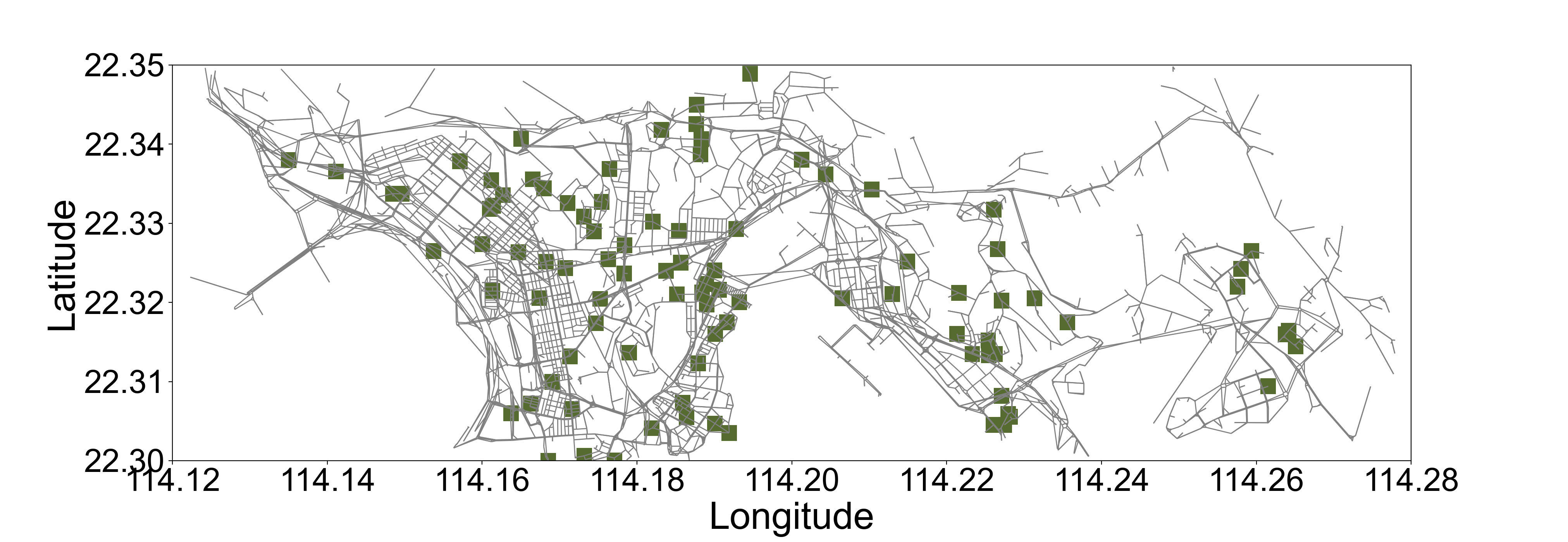}}
    \caption{\textbf{Distributions of requests and vehicles in Hong Kong.} Hong Kong is near the sea and there are many mountains. Thus, the street network topology is extremely complex. However, the distributions of requests as well as vehicles are relatively uniform. The request arrival rate in (a) is 0.49 \#/s, and the number of vehicles in (b) is 100.}
    \label{fig4}
\end{figure}

\begin{figure}[!h]
    \centering
    \subfigure[]{\includegraphics[width=0.51\linewidth]{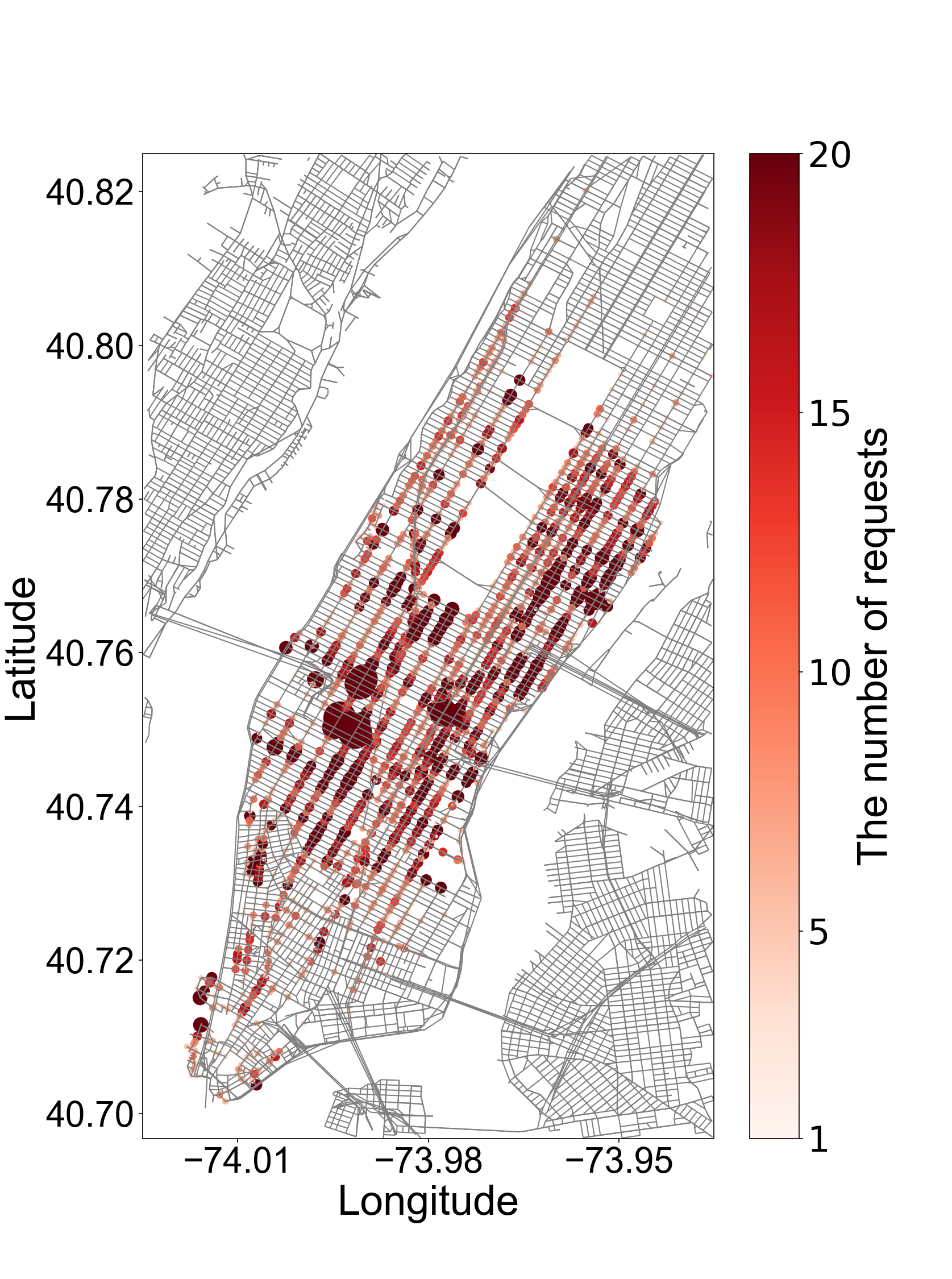}}
    \subfigure[]{\includegraphics[width=0.4\linewidth]{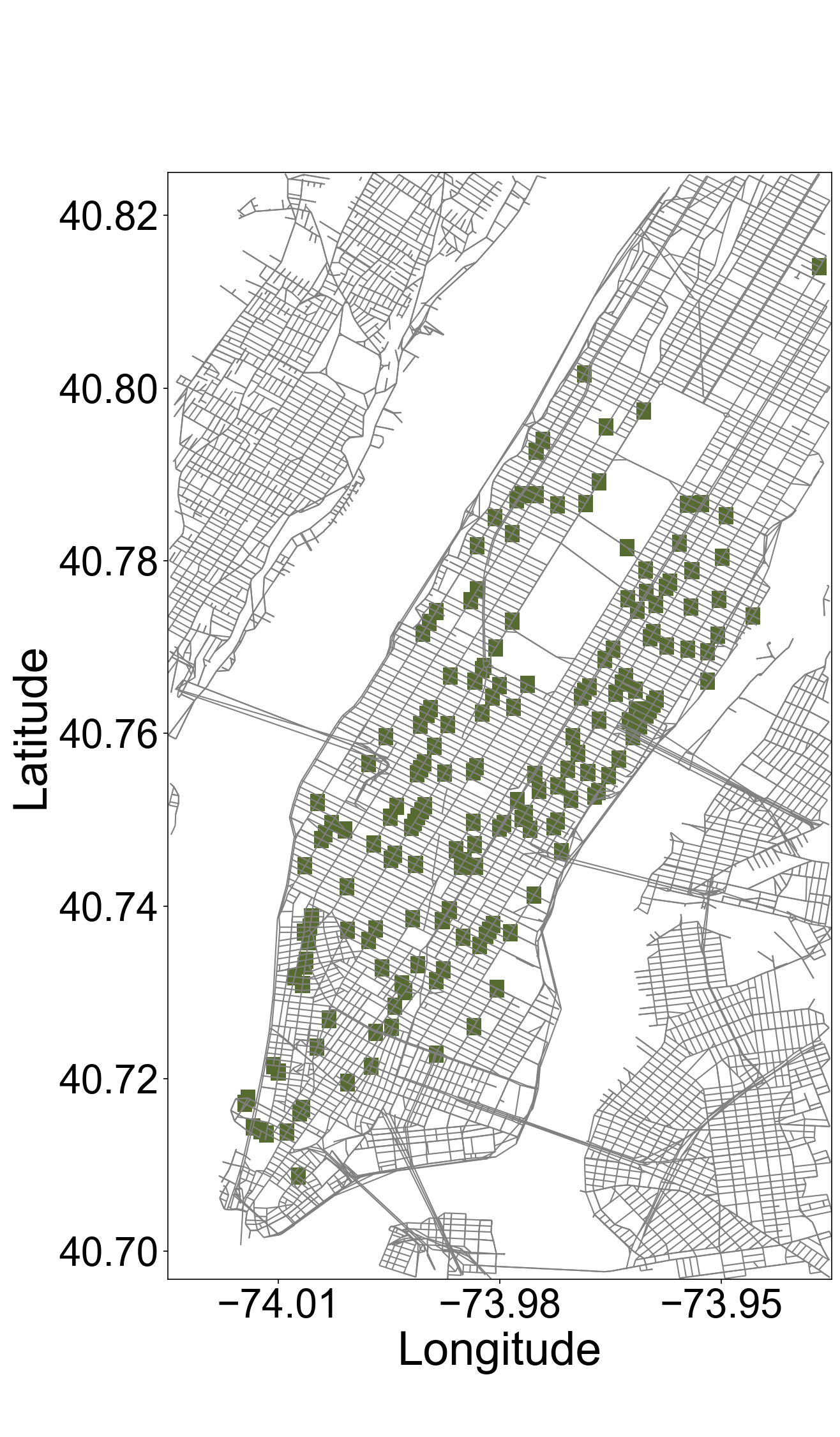}}
    \caption{\textbf{Distributions of requests and vehicles in Manhattan.} Most requests as well as vehicles are concentrated in the south of Central Park. The street network topology is regular, but the study area is banding. The request arrival rate in (a) is 1.42 \#/s, and the number of vehicles in (b) is 200.}
    \label{fig5}
\end{figure}

The datasets used in this study were obtained from the New York Taxi and Limousine Commission (TLC) for July 2015, Hong Kong eTaxi company, and Didi Chuxing for November 2016, for Manhattan, Hong Kong, and Chengdu, respectively. The initial positions of vehicles are generated according to the spatial distributions of trip requests, with hot areas initially having more vehicles. The spatial distributions of requests and vehicles are shown in Figs. \ref{fig3} - \ref{fig5}. Specifically, in Chengdu, the downtown area is the study zone, with the northeastern area being relatively hot in terms of request arrival rate, and correspondingly more vehicles are initialized in this zone (Fig. \ref{fig3}). The street network in Hong Kong is more complex due to its special geography (near the sea and there are many mountains). However, the distributions of requests and the generated vehicles are relatively uniform (Fig. \ref{fig4}). As for Manhattan, most requests are concentrated in the south of Central Park, with most vehicles initialized in this area (Fig. \ref{fig5}). The street network topologies as well as the distributions of requests vary significantly among the three cities, indicating the general applicability of the derived scaling laws based on the experimental results in these cities.

We filter datasets used in this study according to the following procedure. First, only trips where both the origin and destination are located within the study area are considered. Second, trips with a distance of no more than 500 meters are not considered. Third, we relocate the origins and destinations of all trips to the closest intersections. Finally, all trips are rescheduled to the closest matching time points according to the matching time interval (e.g., 2 seconds). To obtain different values of the system load $u$ for a specific fleet size, we uniformly and randomly subsample trips from the original datasets. This results in different request arrival rates $\lambda$ but does not alter the demand patterns.

The street networks including both roads and intersections are obtained from Open Street Map (\cite{OpenStreetMap}). All trips in the datasets consist of time points for sending requests, longitude-latitude pairs of origins and destinations, and the shortest routes from origins to destinations. The shortest routes are estimated using Dijkstra's algorithm (\cite{dijksta1959note}) based on the road networks and are used to calculate the average trip distance. It is worth noting that the travel time is directly computed as distance divided by a constant vehicle speed $v$ since this study does not consider the effect of traffic congestion.

\subsection{Dispatching algorithms}\label{sec: 4.2}

The dispatching algorithm used in this study is similar to the one proposed in \cite{alonso2017demand}. Initially, each trip request is preassigned to vehicles within its matching area. It is important to note that the matching radius is set relatively large to ensure that all trip requests can be perceived by vehicles, even when the vehicle density is low (see Supplementary Information for details). Subsequently, for each vehicle, the pairwise shareability of its potential requests and current request(s) is checked based on pickup time and detour time constraints. This is done to establish a Rider-Trip-Vehicle-graph (RTV-graph), where potential trips and vehicles that can execute them are connected. The optimal assignment is then determined using integer linear programming (ILP). Finally, vehicles conduct the assignment according to their planned routes. Passengers who are not assigned to any vehicles at this step will continue to wait until they are successfully assigned or cancel their orders due to the maximal waiting time constraint. Drivers who do not match any requests will remain where they are. Please refer to the Dispatching Algorithm in the Supplementary Information for more information.

\subsection{Simulations and statistics}\label{sec: 4.3}

To conduct the matching process described above, we develop a simulation platform using Python. The ILP is solved using a commercial solver, CPLEX. At each matching step, we count the average number of scheduled passengers for all vehicles, including passengers already on the vehicle as well as passengers scheduled to be picked up in the future. The average number of scheduled passengers $\Bar{C}$ is calculated as the mean value of the counting results during the entire simulation process. At the end of the simulation, the average service rate $\Bar{R}$ is calculated as the number of served passengers divided by the total number of passengers. It should be noted that we assume each request contains only one passenger, so there is no difference between a trip request and a passenger in this study.

\section*{Data availability} 
The Manhattan dataset is provided by New York TLC and cab be available at \url{https://www1.nyc.gov/site/tlc/about/tlc-trip-record-data.page}. The Hong Kong and Chengdu datasets are provided by a taxi operator in Hong Kong and Didi Chuxing respectively; they are not publicly available due to privacy protection issues.

\section*{Code availability}
The code is available at \url{https://github.com/HKU-Smart-Mobility-Lab/Ride-sharing-Simulator}.

\newpage

\bibliography{reference}

\newpage

\section*{Acknowledgements}
The work described in this paper was partly supported by a smart traffic fund (PSRI/29/2201/PR) jointly funded by the Hong Kong Productivity Council and Transport Department of Hong Kong SAR Government. It was also partially supported by a research grant from Hong Kong Research Grant Council (RGC) under a general research grant (GRF) HKU15209121. The authors also thank Mr. Taijie Chen and Mr. Yuhao Zhang for pre-processing the data.

\section*{Author contributions}
Wang Chen: Methodology, Software, Formal analysis, Writing - original draft. Jintao Ke: Conceptualization, Supervision, Writing - review \& editing. Linchuan Yang: Conceptualization, Writing - review \& editing.

\section*{Competing interests}
The authors declare no competing interests.

\newpage

\newpage

\newpage

\newpage

\newpage

\newpage
\begingroup
\setlength{\tabcolsep}{10pt} 
\renewcommand{\arraystretch}{1} 
\begin{table}[htbp!]
\caption{Error measures of $\Bar{C}$}
\begin{center}
\begin{tabular}{ c|c|c|c|c|c|c } 
 \hline
 City & Maximal capacity & R$^2$ & MSE & RMSE & MAE & MAPE (\%) \\
 \hline
Chengdu     & C=2 & 0.991 & 0.002 & 0.039 & 0.033 & 3.2 \\
            & C=3 & 0.988 & 0.005 & 0.065 & 0.053 & 3.8 \\
            & C=4 & 0.969 & 0.017 & 0.121 & 0.099 & 5.4 \\
            & C=6 & 0.889 & 0.093 & 0.283 & 0.223 & 9.3 \\
Hong Kong   & C=2 & 0.936 & 0.016 & 0.126 & 0.112 & 7.0 \\
            & C=3 & 0.954 & 0.023 & 0.152 & 0.135 & 6.6 \\
            & C=4 & 0.985 & 0.011 & 0.102 & 0.087 & 3.9 \\
            & C=6 & 0.98 & 0.024 & 0.145 & 0.111 & 3.9 \\
Manhattan   & C=2 & 0.889 & 0.023 & 0.151 & 0.137 & 7.9 \\
            & C=3 & 0.885 & 0.058 & 0.235 & 0.214 & 9.2 \\
            & C=4 & 0.925 & 0.065 & 0.242 & 0.215 & 7.9 \\
            & C=6 & 0.971 & 0.043 & 0.201 & 0.162 & 5.1 \\
\hline
\end{tabular}
\label{table1}
\end{center}
\end{table}

\newpage
\begingroup
\setlength{\tabcolsep}{10pt} 
\renewcommand{\arraystretch}{1} 
\begin{table}[htbp!]
\caption{Error measures of $\Bar{R}$}
\begin{center}
\begin{tabular}{ c|c|c|c|c|c|c } 
 \hline
 City & Maximal capacity & R$^2$ & MSE & RMSE & MAE & MAPE (\%) \\
 \hline
Chengdu     & C=2 & 0.996 & 0.000 & 0.015 & 0.010 & 1.5 \\
            & C=3 & 0.984 & 0.001 & 0.028 & 0.021 & 3.1 \\
            & C=4 & 0.980 & 0.001 & 0.031 & 0.024 & 3.9 \\
            & C=6 & 0.957 & 0.002 & 0.042 & 0.036 & 6.7 \\
Hong Kong   & C=2 & 0.976 & 0.001 & 0.037 & 0.034 & 6.7 \\
            & C=3 & 0.963 & 0.002 & 0.044 & 0.038 & 6.7 \\
            & C=4 & 0.977 & 0.001 & 0.033 & 0.025 & 3.9 \\
            & C=6 & 0.984 & 0.001 & 0.028 & 0.024 & 4.3 \\
Manhattan   & C=2 & 0.960 & 0.003 & 0.054 & 0.041 & 8.6 \\
            & C=3 & 0.933 & 0.005 & 0.066 & 0.052 & 9.6 \\
            & C=4 & 0.932 & 0.004 & 0.063 & 0.049 & 8.3 \\
            & C=6 & 0.959 & 0.002 & 0.044 & 0.033 & 5.2 \\
\hline
\end{tabular}
\label{table2}
\end{center}
\end{table}

\newpage

\begin{center}
    \begin{tabular}{c}
     \\
     \\
     \\
     \\
     \\
     \\
     \\
     \\
     \textbf{Supplementary Information}  \\
     
     \\Scaling Laws of Dynamic High-Capacity Ride-Sharing \\

     \\Wang Chen$^a$, Jintao Ke$^a$, Linchuan Yang$^b$ \\

     \\$^a$ Department of Civil Engineering,\\
     the University of Hong Kong, Hong Kong, China
     \\$^b$ Department of Urban and Rural Planning, School of Architecture and Design,\\
     Southwest Jiaotong University, Chengdu, China\\
     
    \end{tabular}
\end{center}

\setcounter{equation}{0}
\setcounter{figure}{0}
\setcounter{table}{0}
\setcounter{section}{0}
\renewcommand\theequation{S.\arabic{equation}}
\renewcommand\thefigure{S\arabic{figure}}
\renewcommand\thetable{S\arabic{table}}
\renewcommand\thesection{S\arabic{section}}

\newpage
\begin{figure}[!h]
    \centering
    \subfigure[Chengdu]{\includegraphics[height = 0.25\linewidth, width=1.0\linewidth]{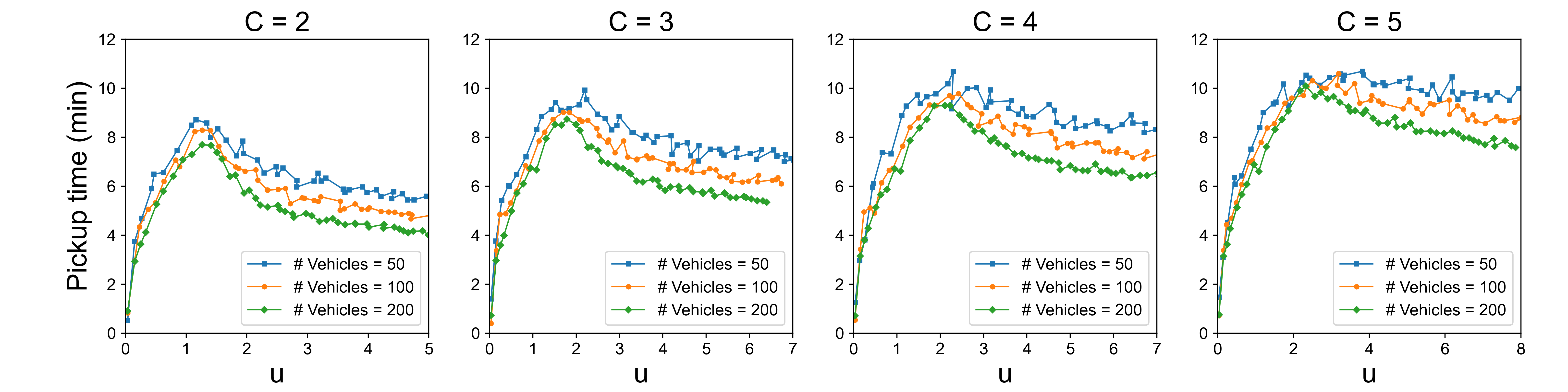}}
    \subfigure[Hong Kong]{\includegraphics[height = 0.25\linewidth, width=1.0\linewidth]{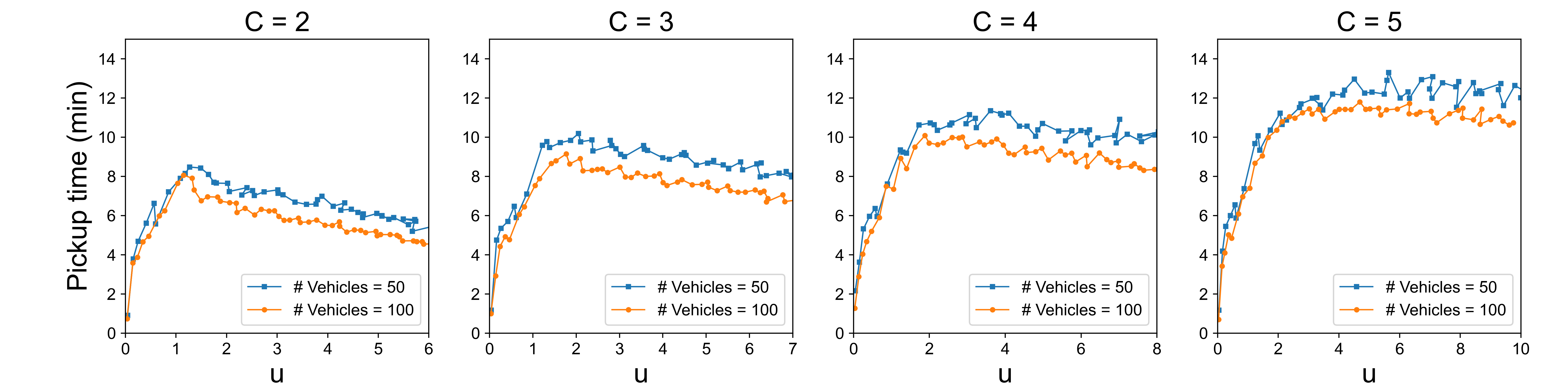}}
    \subfigure[Manhattan]{\includegraphics[height = 0.25\linewidth, width=1.0\linewidth]{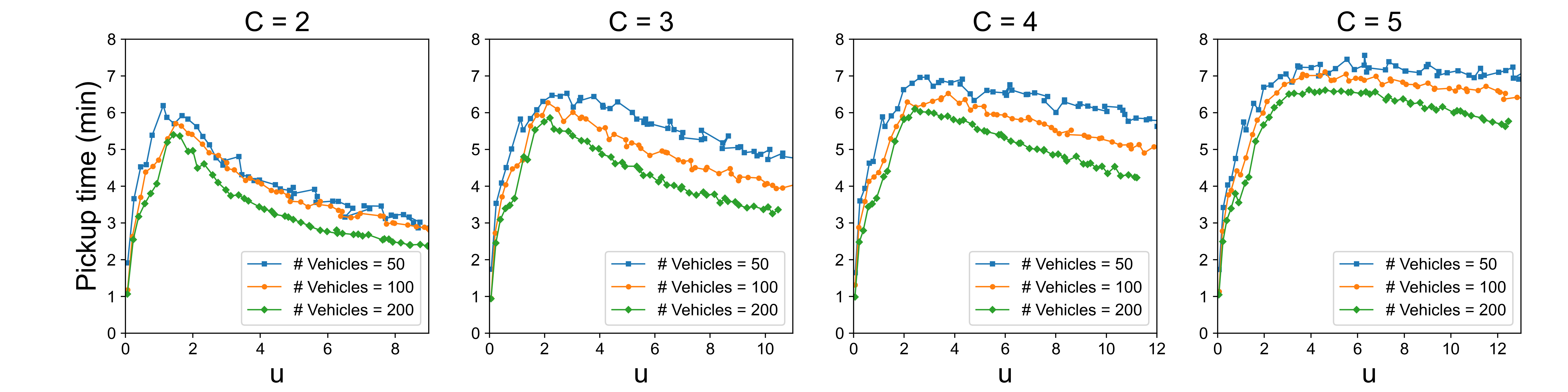}}
    \caption{\textbf{Experimental results of pickup time.} The maximal pickup time is no more than 15 minutes. This verifies the reasonability of the assumption that the matching radius can be relatively large.}
    \label{figs0}
\end{figure}

\newpage
\begin{figure}[!h]
    \centering
    \subfigure[Chengdu]{\includegraphics[height = 0.25\linewidth, width=1.0\linewidth]{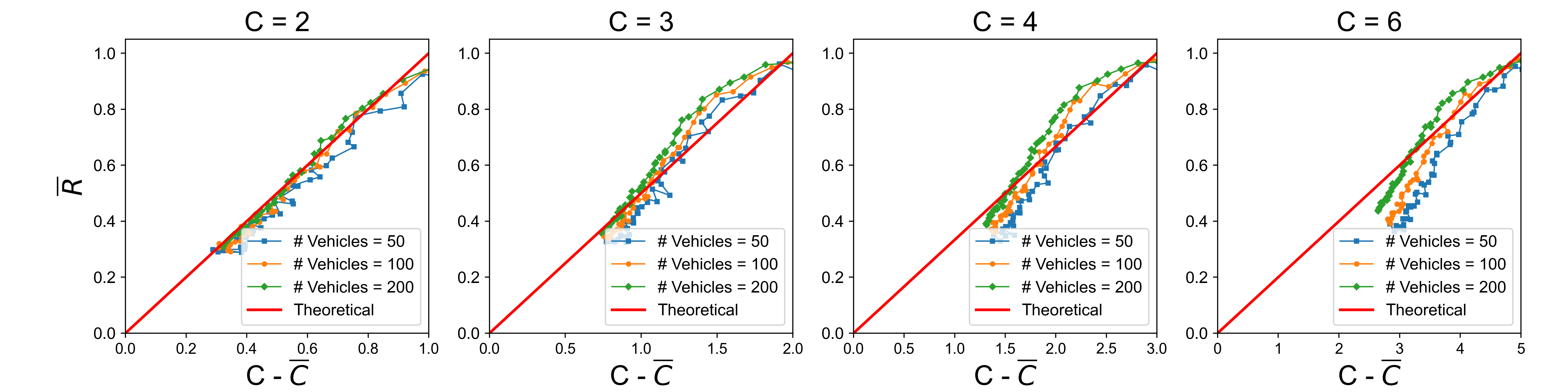}}
    \subfigure[Hong Kong]{\includegraphics[height = 0.25\linewidth, width=1.0\linewidth]{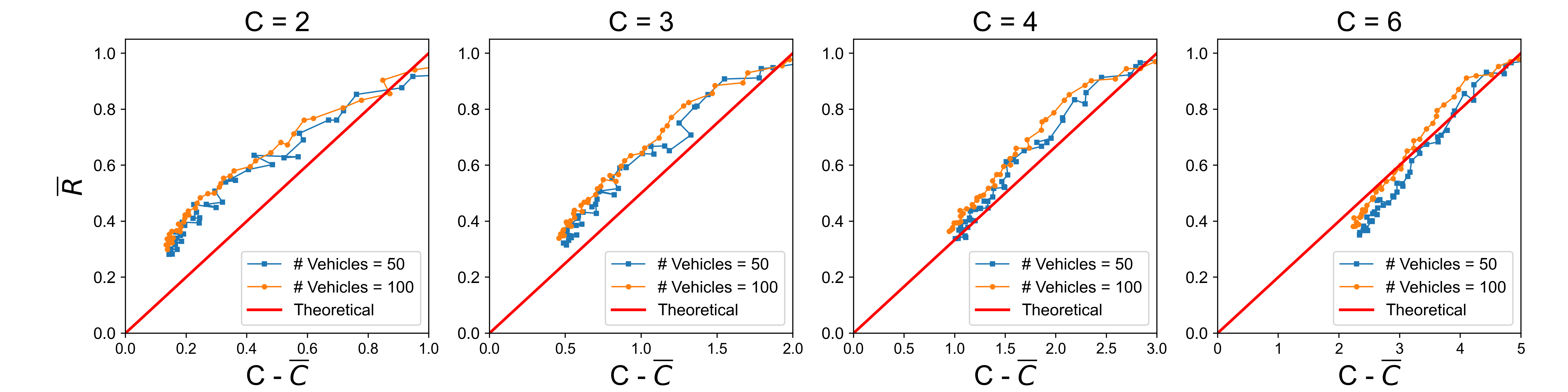}}
    \subfigure[Manhattan]{\includegraphics[height = 0.25\linewidth, width=1.0\linewidth]{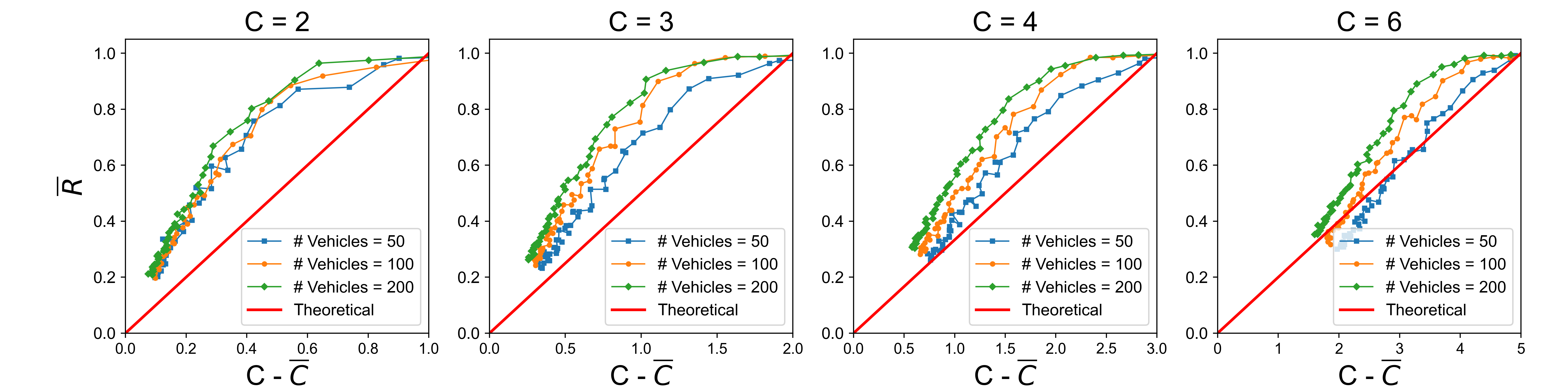}}
    \caption{\textbf{Experimental and theoretical curves of $\Bar{R}$ with respect to $C-\Bar{C}$.} Experimental results in three cities (Chengdu, Hong Kong, and Manhattan) demonstrate that the correlation between the average service rate of passengers $\Bar{R}$ and the remaining capacity of vehicles $C-\Bar{C}$ is approximately linear. The theoretical curves are derived as $\Bar{R} = \frac{1}{C-1}(C - \Bar{C})$.}
    \label{figs1}
\end{figure}

\newpage
\begin{figure}[!h]
    \centering
    \setcounter{subfigure}{0}
    \subfigure[Chengdu]{\includegraphics[height = 0.25\linewidth, width=1.0\linewidth]{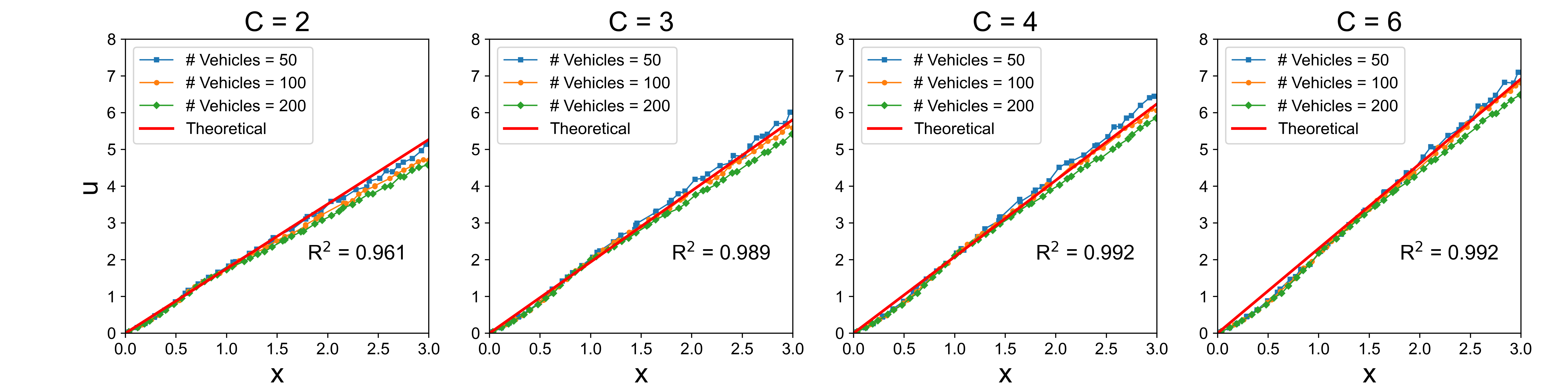}}
    \subfigure[Hong Kong]{\includegraphics[height = 0.25\linewidth, width=1.0\linewidth]{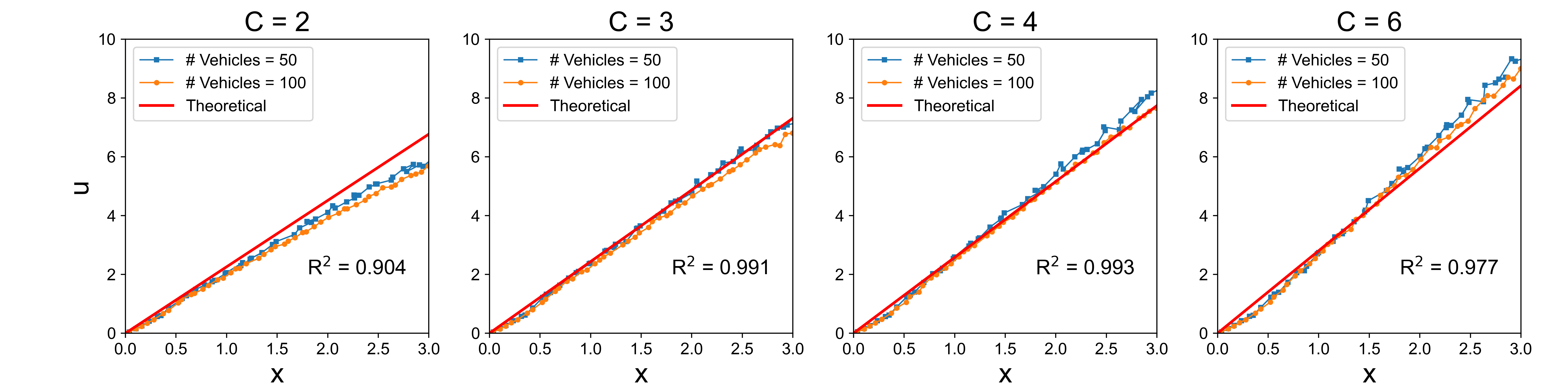}}
    \subfigure[Manhattan]{\includegraphics[height = 0.25\linewidth, width=1.0\linewidth]{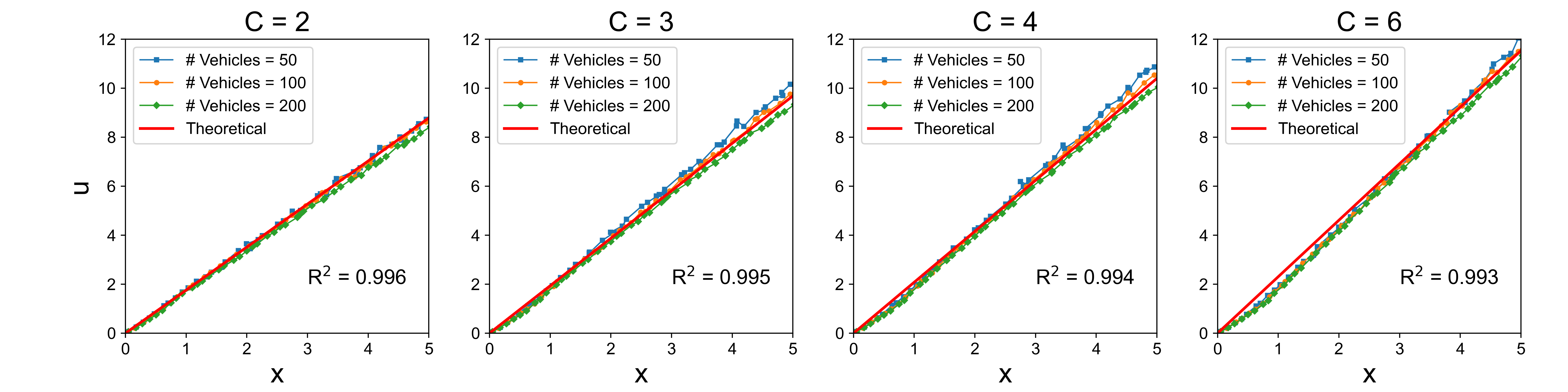}}
    \caption{\textbf{Approximation of $u$.} The system load $u$ is directly proportional to normalized load $x$, with $R^2$ greater than 0.9 across all scenarios. The proportional ratio depends on the maximal allowed detour time ratio, the complexity of road networks, and maximal vehicle capacities.}
    \label{figs3}
\end{figure}

\newpage
\begin{figure}[!h]
    \centering
    \includegraphics[width=1.0\linewidth]{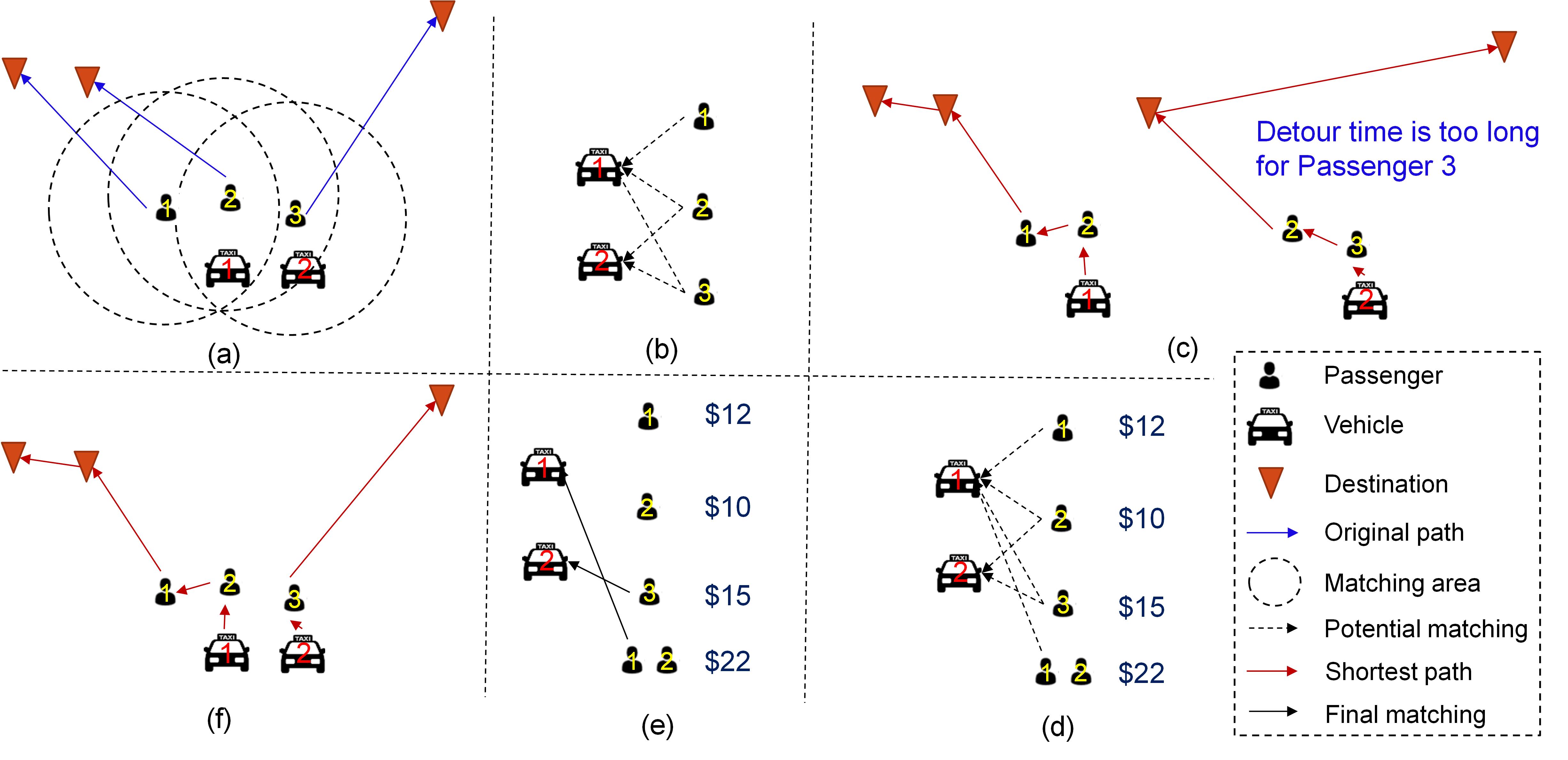}
    \caption{\textbf{Dispatching Algorithm.} (a) Passengers are preassigned to vehicles within their matching areas; then (b) each vehicle will be potentially scheduled with multiple passengers. (c) The platform checks the shareability of each vehicle's potential passengers by planning the shortest routes and verifying pickup and detour time constraints, e.g., Passengers 1 and 2 can share Vehicle 1, but Passengers 2 and 3 cannot share Vehicle 2 due to the detour time constraint. (d) RTV-graph can be established to connect all potential trips (including one or more requests) to vehicles. (e) The optimal matching results are obtained via ILP. (f) Vehicles pick up and deliver passengers according to the matching results of (e) and the planned shortest routes of (c).}
    \label{figS4:DispatchingAlg}
\end{figure}

\newpage
\section{Derivation}\label{derivation}

In a stationary state, the average number of scheduled passengers of each vehicle $\Bar{C}$ is directly related to the average service time $\Bar{t}$ (including pickup time and delivery time) and the average service rate $\Bar{R}$ (\cite{molkenthin2020scaling}). During the average service time $\Bar{t}$ of a request, on average, a total of $\lambda\Bar{t}$ new requests join the ride-sharing system. Therefore, $\lambda\Bar{t}\Bar{R}$ requests are scheduled. The average scheduled requests of each vehicle is $\lambda\Bar{t}\Bar{R}/N$. In a stationary state, new requests can only be served after older ones have been delivered. Thus, the average number of scheduled passengers $\Bar{C}$ is equal to the average scheduled requests of each vehicle during the average service time:

\begin{equation}
    \label{eq:c_bar_}
    \Bar{C} = \frac{\lambda\Bar{t}\Bar{R}}{N}
\end{equation}

For the average service rate of passengers $\Bar{R}$, in a stationary state, it can be represented as the matching probability of a new request joining the ride-sharing system within the time of $1/\lambda$. To estimate the matching probability, we consider two scenarios. First, when less than 1 passenger is scheduled per vehicle on average, i.e., $\Bar{C} \leq 1$, there exists at least one idle vehicle. Therefore, if the matching radius is large enough, a new passenger can be immediately scheduled to the idle vehicle once he/she sends a request to the ride-sharing system. The matching probability as well as the average service rate $\Bar{R}$, therefore, are equal to 1:

\begin{equation}
    \label{eq:R_bar=1}
    \Bar{R} = 1
\end{equation}
The matching radius, however, cannot be infinite because it is not practical for passengers to wait for a driver who is dozens of kilometers away from them to pick them up. However, in a practical dynamic ride-sharing system, the distribution of vehicles is expected to be relatively dense. This ensures that Eq. \eqref{eq:R_bar=1} holds when the matching radius is relatively large. The experimental results show that the maximal pickup time is no more than 15 minutes (Fig. \ref{figs0}), which verifies the reasonability of Eq. \eqref{eq:R_bar=1}.

On the other hand, when each vehicle has been scheduled for at least one passenger on average, i.e., $\Bar{C} > 1$, ride-sharing should be considered. It is significantly difficult to estimate high-capacity ride-sharing probabilities through mathematically modeling features of the ride-sharing system like \cite{tachet2017scaling}. Instead, we infer the matching probability of a new request based on physical properties. Intuitively, the more seats of vehicles left, the higher the matching probability of requests. In other words, no matter how many passengers can be scheduled or have been scheduled per vehicle, the matching probability of a new request is positively correlated with the average remaining capacity of vehicles, i.e., $(C-\Bar{C})$. To quantify the correlation, we plot the average service rate $\Bar{R}$ as a function of the remaining capacity $(C-\Bar{C})$ based on the experimental results. As shown in Fig. \ref{figs1}, the correlation is approximately linear. Therefore, we infer the general matching probability as well as the average service rate $\Bar{R}$ as follows:

\begin{equation}
    \label{eq:R_bar=k(C_bar-C)}
    \Bar{R} = k_{c}(C-\Bar{C})
\end{equation}
where $k_{c}$ denotes the correlation factor. It should be noted that $k_{c}$ is correlated with the maximal capacity $C$.

Now, we can solve Eqs.~\eqref{eq:c_bar_}-\eqref{eq:R_bar=k(C_bar-C)} simultaneously to obtain the representations of $\Bar{C}$ and $\Bar{R}$:

\begin{equation}
    \label{eq:C_bar_SI}
    \Bar{C}=\left\{
        \begin{array}{cl}
            u         & u \leq u_{1} \\
            \frac{u}{\frac{1}{k_{c}}+u} \cdot C & u > u_{1}   \\
        \end{array} \right.
\end{equation}

\begin{equation}
    \label{eq:R_bar_SI}
    \Bar{R}=\left\{
        \begin{array}{cl}
            1         &  u \leq u_{1} \\
            \frac{1}{\frac{1}{k_{c}} + u} \cdot C & u > u_{1}   \\
        \end{array} \right.
\end{equation}
where $u_{1}$ denotes the point where $\Bar{C}=1$. The function of $\Bar{C}$ is supposed to be continuous, thus we have:

\begin{equation}
    \label{eq:C_bar_continuous}
    u_{1}=1=\frac{u_{1}}{\frac{1}{k_{c}}+u_{1}} \cdot C
\end{equation}
the $k_{c}$ and $u_{1}$ can be derived as:

\begin{equation}
    \label{eq:kc}
    k_{c}=\frac{1}{C-1}
\end{equation}
\begin{equation}
    \label{eq:u1}
    u_{1}=1
\end{equation}
Actually, we can obtain the same result by using the continuity of $\Bar{R}$. It should be noted that $k_{c}$ is a constant related to $C$, which further verifies the linear relationship between $\Bar{R}$ and the remaining capacity $(C-\Bar{C})$ when $\Bar{C} > 1$. By substituting Eqs. \eqref{eq:kc} and \eqref{eq:u1} into Eqs. \eqref{eq:C_bar_SI}-\eqref{eq:R_bar_SI}, we can obtain the scaling laws listed in Eqs. \eqref{eq:C_bar}-\eqref{eq:R_bar}.

\section{Approximation of $u$}\label{approximate_u}

It is difficult to accurately estimate the average service time of each passenger due to the complex street network topologies, the ever-changing spatial-temporal distributions of vehicles and passengers, and the number of scheduled passengers. Therefore, we attempt to approximate the system load $u$ using a few exogenous variables predefined in the dynamic ride-sharing system, such as the maximal capacity of vehicles $C$.

\cite{molkenthin2020scaling} introduced a dimensionless parameter $x$ named normalized load as follows:

\begin{equation}
    \label{x}
    x = \frac{\lambda\Bar{d}}{Nv}
\end{equation}
Remind that the system load $u$ is defined as $u = \frac{\lambda}{N/\Bar{t}}$. The two formulas are quite similar, but the key distinction is that the normalized load $x$ does not contain any endogenous variables, allowing $x$ can be calculated once the ride-sharing system is designed or observed. Therefore, if the system load $u$ can be represented by the normalized load $x$, we can approximate $u$ using a few exogenous variables.

When plotting the experimental curves of $u$ with respect to $x$ (as shown in Fig. \ref{figs3}), we observe that $u$ is directly proportional to $x$ and the proportional ratio depends on a few exogenous variables, i.e., the maximal capacity of vehicles $C$, the maximal allowed detour time ratio $r_{dt}$, and the complexity of the road network $T$. Specifically, $r_{dt}$ represents the maximal allowable ratio of detour time to the original trip time (in-vehicle time) without any detours. For example, if the original trip time of a request is 10 minutes and $r_{dt}$ is 0.5, then the maximal allowable detour time is $0.5 \times 10 = 5$ minutes, resulting in the maximal allowable delivery time is $5 + 10 = 15$ minutes. The complexity of the road network $T$ is used to capture the road network topology. A more complex network topology corresponds to a higher value of $T$, as well as more detour time required to pick up and deliver shared trips. In this study, the network topologies in Chengdu and Manhattan are relatively regular, with a value of $T$ equaling to 0. In contrast, the network topology in Hong Kong is significantly more complex, with a $T$ value of 0.5.

We now approximate $u$ with respect to $x$ as follows:

\begin{equation}
    \label{eq:u_x}
    u = (r_{dt} + T + \sqrt[3]{C})x
\end{equation}
As shown in Fig. \ref{figs3}, the proposed approximation formula of $u$ has a strong agreement with experimental results, with $R^2$ more than 0.9 across all scenarios. This demonstrates that $u$ can be accurately estimated by $x$ and a few exogenous variables. In addition, it is worth noting that all terms in the formula for $u$ can be calculated once the dynamic ride-sharing system is designed or observed. This makes it easy to approximate $u$. Since $u$ is the only parameter that can be accurately and easily approximated, the proposed scaling laws in this study are general enough to be extrapolated to other cities.


\section{Dispatching Algorithm}\label{Dispatching Algorithm}

As shown in Fig. \ref{figS4:DispatchingAlg}, the dispatching process is conducted step by step as follows. First, the ride-sharing platform collects all requests $R$ within a given time interval (e.g., 2 seconds) and pre-assigns each passenger to vehicles within their matching area (Fig. \ref{figS4:DispatchingAlg} (a)). Thus, each of the vehicles $J$ may be scheduled to pick up and deliver multiple passengers (Fig. \ref{figS4:DispatchingAlg} (b)). It should be noted that if a passenger's matching area has no available vehicle, the platform will not process their request during the current step. Subsequently, the platform determines which two or more passengers can be paired via planning the shortest routes and verifying pickup and detour time constraints. Specifically, as shown in Fig. \ref{figS4:DispatchingAlg} (c), the shortest route for Vehicle 2 to pick up and deliver Passenger 2 and 3 is indicated by the red arrows. The total trip time for Passenger 3 is approximately twice as long as original trip time (without ride-sharing), meaning the detour time ratio $t_{d}$ is nearly 1 that is greater than the constraint (i.e., 0.5). Therefore, Passenger 2 and 3 cannot share Vehicle 2. The shortest or optimal routes can be estimated by enumeration or heuristic algorithms. For more details about these algorithms, we refer readers to \cite{alonso2017demand}. Once all ride-sharing trips (each trip includes one or more requests) have been determined, the RTV-graph can be established to connect potential trips $I$ and vehicles $J$ (Fig. \ref{figS4:DispatchingAlg} (d)). Then, the platform assigns trips $I$ to available vehicles $J$ via ILP by determining the binary variable $x_{ij}$ for all $i\in I$ and $j\in J$ to maximize the expected total value:

\begin{equation}
    \label{eq:ILP}
    \max_{x}{\sum_{i\in I}\sum_{j\in J}{x_{ij}\cdot v_{ij}}}
\end{equation}
s.t.

\begin{equation}
    \label{eq:con1}
    \sum_{i\in I}{x_{ij}\leq 1, \quad \forall j\in J}
\end{equation}

\begin{equation}
    \label{eq:con2}
    \sum_{j\in J}\sum_{r\in i; i\in I}{x_{ij}\leq 1, \quad \forall r\in R}
\end{equation}

\begin{equation}
    \label{eq:xij}
    x_{ij}\in \{ 0,1 \} \quad \forall i\in I, j\in J
\end{equation}
where $x_{ij} = 1$ denotes that trip $i$ is scheduled to vehicle $j$. Constraints \eqref{eq:con1} ensure that each vehicle can be assigned at most one trip, and constraints \eqref{eq:con2} ensure that each request can be scheduled to at most one vehicle. $v_{ij}$ denotes the value for vehicle $j$ to pick up and deliver trip $i$ and can be expressed as follows:

\begin{equation}
    \label{eq:vij}
    v_{ij} = \sum_{r\in i}{(p_r - c_{rj})}
\end{equation}
where $p_{r}$ denotes the value of request $r$, $c_{rj}$ the cost for vehicle $j$ to pick up and deliver request $r$. In this study, the values of all requests are identical and the cost is calculated as delayed time including pickup time and detour time.The resulting optimization problem can be solved using commercial optimization solvers such as CPLEX. The optimization results are the final assignments that determine which vehicle will pick up and deliver which trip (Fig. \ref{figS4:DispatchingAlg} (e)). Finally, vehicles execute the assignments according to the planned shortest routes (Fig. \ref{figS4:DispatchingAlg} (f)). Vehicles that have no assignments in this step will remain in their current location, and unscheduled requests will wait for the next matching step until they leave the system due to the maximal waiting time constraint.
\\

\section{Error Measurement Metrics}\label{Error}

We adopt R square (R$^2$), Mean Square Error (MSE), Root Mean Square Error (RMSE), Mean Absolute Error (MAE), and Mean Absolute Percentage Error (MAPE) to measure the errors between the experimental results $y$ and the prediction of derived theoretical models $\hat{y}$. The five metrics are calculated as follows:

\begin{equation}
    \label{R2}
    R^2 = \frac{1}{F}\sum_{f=1}^{F}(1 - \frac{\sum_{i=1}^{N}(y_{i}^{f} - \hat{y}_{i}^{f})^{2}}{\sum_{i=1}^{N}(y_{i}^{f} - \frac{1}{N}\sum_{i=1}^{N}y_{i}^{f})^{2}})
\end{equation}

\begin{equation}
    \label{MSE}
    MSE = \frac{1}{F\cdot N}\sum_{f=1}^{F}{\sum_{i=1}^{N}(y_{i}^{f} - \hat{y}_{i}^{f})^{2}}
\end{equation}

\begin{equation}
    \label{RMSE}
    RMSE = \sqrt{\frac{1}{F\cdot N}\sum_{f=1}^{F}{\sum_{i=1}^{N}(y_{i}^{f} - \hat{y}_{i}^{f})^2}}
\end{equation}

\begin{equation}
    \label{MSE}
    MAE = \frac{1}{F\cdot N}\sum_{f=1}^{F}{\sum_{i=1}^{N}\lvert y_{i}^{f} - \hat{y}_{i}^{f}\rvert}
\end{equation}

\begin{equation}
    \label{MSE}
    MAPE = \frac{1}{F\cdot N}\sum_{f=1}^{F}{\sum_{i=1}^{N}\lvert \frac{y_{i}^{f} - \hat{y}_{i}^{f}}{y_{i}^{f}}\rvert}
\end{equation}
where $F$ denotes the number of scenarios with different fleet sizes ($F=3$ for Chengdu and Manhattan and $F=2$ for Hong Kong in this study), and $N$ the number of samples in each scenario.

\end{document}